\documentclass{llncs}

\usepackage{natbib}
\usepackage{graphicx}

\usepackage{fancyhdr}
\usepackage{lastpage}
 
\begin{document}

\title{A short review and primer on multimodal psychophysiological applications in work-related human computer interaction}
\author{Marco Filetti\inst{1}, Jari Torniainen\inst{2}}
\institute{Helsinki Institute for Information Technology HIIT, Department of Computer Science, Aalto University, Espoo, Finland\\
\email{marco.filetti@aalto.fi}
\and
Quantitative Employee unit, Finnish Institute of Occupational Health, Helsinki, Finland}

\maketitle              

\begin{abstract}

The application of psychophysiology in human-computer interaction is a growing field with significant potential for future smart personalised systems. Working in this emerging field requires comprehension of an array of physiological signals and analysis techniques. 

This paper focuses on the aggregation of multiple physiological measurements, obtained from one or more sensors. This approach requires the classification of relatively large samples of multidimensional data, which must be associated to specific cognitive or affective states. Researchers generally attempt to solve this problem by utilising machine learning techniques. 

We present a short review to serve as a primer for the novice, enabling rapid familiarisation with the latest core concepts. We put special emphasis on work-related human-computer interface applications to distinguish from the more common clinical or sports uses of psychophysiology.

This paper is an extract from a comprehensive review of the entire field of ambulatory psychophysiology, including 12 similar chapters, plus application guidelines and systematic review. Thus any citation should be made using the following reference:

{\parshape 1 2cm \dimexpr\linewidth-1cm\relax
B. Cowley, M. Filetti, K. Lukander, J. Torniainen, A. Henelius, L. Ahonen, O. Barral, I. Kosunen, T. Valtonen, M. Huotilainen, N. Ravaja, G. Jacucci. \textit{The Psychophysiology Primer: a guide to methods and a broad review with a focus on human-computer interaction}. Foundations and Trends in Human-Computer Interaction, vol. 9, no. 3-4, pp. 150--307, 2016.
\par}


\end{abstract}

\section{Introduction}

\begin{figure}[t]
   \centering
   \includegraphics[scale=0.40]{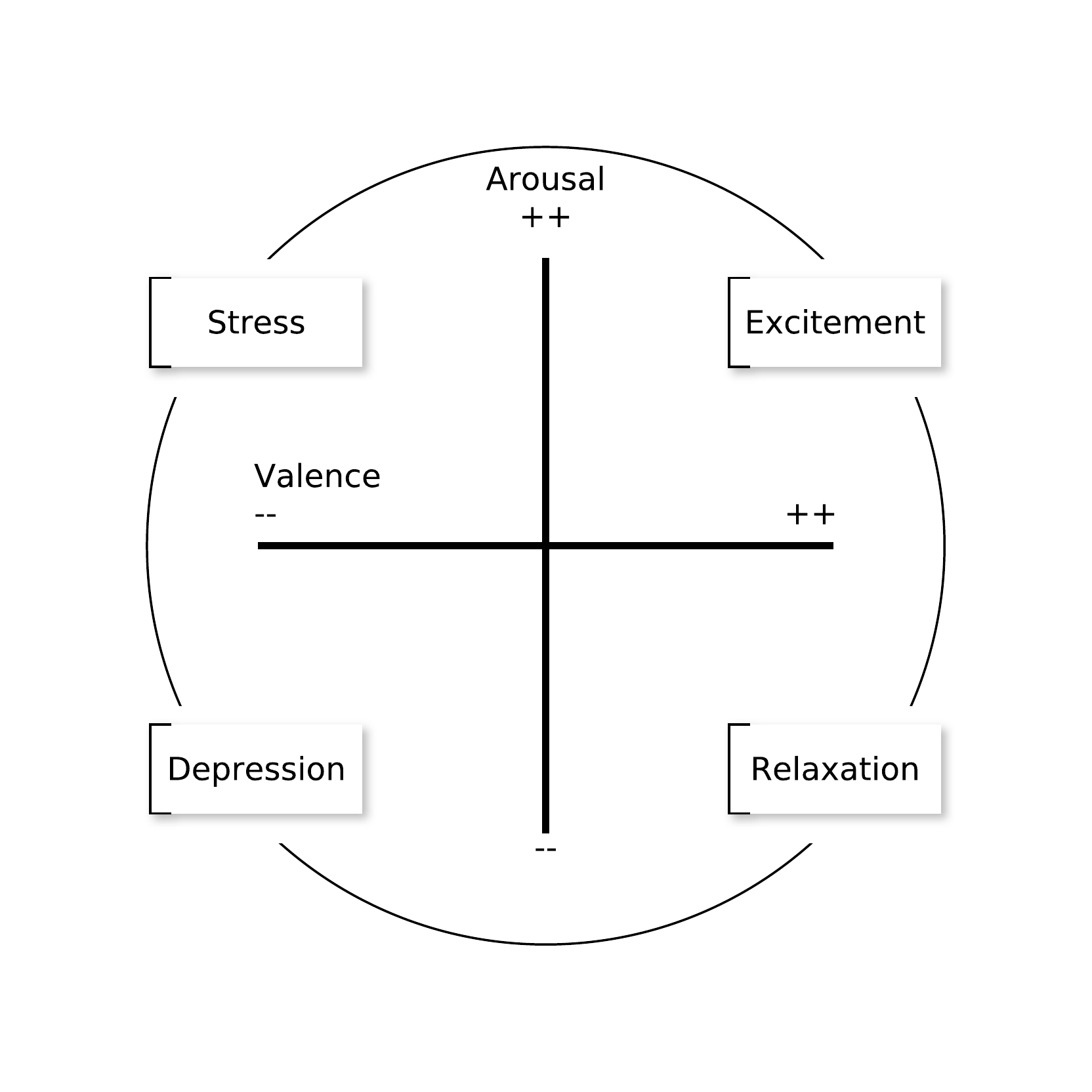}
   \caption{The simple emotional circumplex model, with orthogonal bipolar dimensions of arousal (from alert to lethargic) and valence (pleasant to unpleasant).}
   \label{fig.emocir}
\end{figure}

Signals obtained from multiple sources (physiological and behavioural) can be combined and analysed
collectively for determining the state of a user (or, in some cases, multiple users). Multiple metrics can also be computed from
a single signal. Using multiple measurements has the potential of providing more accurate estimates:
while data collected from a single source could lead to
conflicting interpretations, using data from multiple sources in combination aids in disambiguating
cognitive or affective state. In the case of the widely known two-dimensional model of affect 
\citep{russell1980circumplex} (a simplified version of which is depicted in Figure~\ref{fig.emocir}), a moderate increase in arousal
could be caused by the subject experiencing either fear or annoyance. For the purpose of our example, we assume that this increase
in arousal is unequivocally measured via EDA (discussed in \cite[Section~3.2]{cowley2016primer}).
Integrating EDA with additional measurement (such as facial recognition) can assist in
discriminating which of the two emotions the user was actually experiencing; for instance, the vertical position of the eyebrows
is often higher when one is experiencing fear rather than annoyance.

Similarly, when facial recognition alone does not provide enough information to detect
the emotion unambiguously, measuring arousal (from EDA or other signals) could aid in correctly 
classifying less clear-cut cases. For example, \citet{bailenson2008real} found males
to be less facially expressive than females when experiencing sadness. Accordingly, including additional physiological
signals in their classifier (such as EDA and ECG) increased the probability of correctly detecting sadness for males. \citet{Mandryk2007} presented an interesting approach, using a fuzzy classifier based on ECG, EDA, and facial EMG, to classify both arousal and valence of participants playing a popular computer game.

\begin{figure}[!ht]
   \centering
   \includegraphics[scale=1.0]{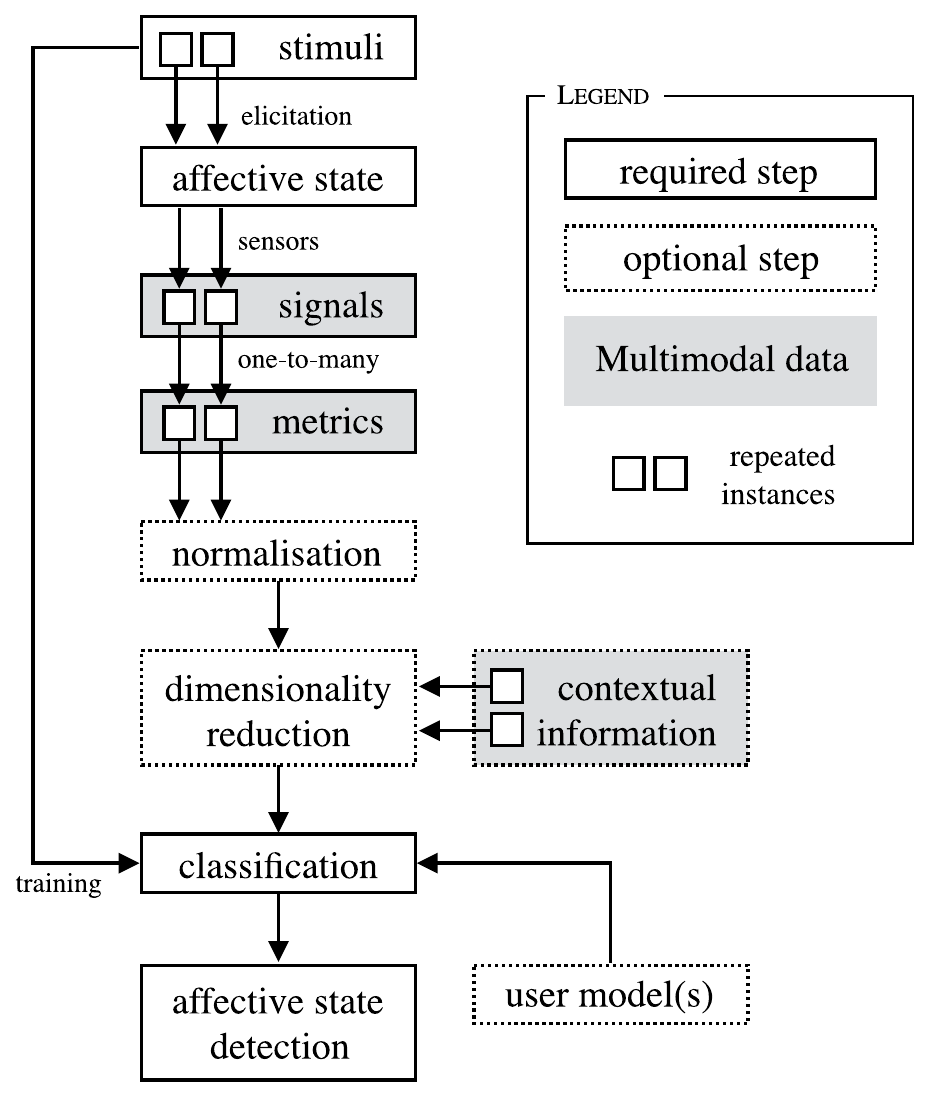}
   \caption{Diagrammatic generalisation of multimodal physiological systems. Grey boxes are roughly equivalent to the concept of a `complex topography' in signal processing systems (as described in \cite[Section~3.12]{cowley2016primer}).}
   \label{fig.multimodal_flowchart}
\end{figure}

\citet{novak2012survey} and \citet{lisetti2004using} discuss general strategies that can be employed in the design and implementation of multimodal systems.
Inspired by these works, we describe here a generic schema for detecting affective and cognitive state in the types of multimodal systems discussed in this part of the primer; see Figure~\ref{fig.multimodal_flowchart}. As this figure shows, stimuli are associated with some cognitive/affective state(s) in the first step, either through the use of validated stimuli (e.g., the International Affective Picture System) or via subjective labelling of each stimulus after perception (for instance, a questionnaire can be used to evaluate a recording of the experience). Metrics are computed from their respective signals. Then, a classifier is trained to associate the data obtained with specific affective states on the basis of the stimulus classification performed earlier. `Optional' procedures, shown in dotted boxes in the figure, may be applied at this stage. Such optional procedures may increase classification accuracy. For example, one may include a user model based on population-level inferences or (non-physiological) contextual information -- such as time of day. Finally, reverse inference is performed to associate a particular set of data with a single specific state or with a range of states; probabilities are used to identify those states that are more likely to be associated with the observed data.

Multimodal systems can be useful when data from
a given channel are missing, a situation that can normally be expected to arise when data are recorded
outside research laboratories. For example, in the research by \citet{wagner2011exploring}, missing data were handled by means of a na\"ive Bayes classifier and three modalities: voice (audio), facial recognition (video), and gestures (accelerometer).

\subsubsection{Determination of affective and cognitive state}
Several studies have considered fusing information from multiple
signals in order to assess mental workload. In a recent
study, \citet{hogervorst:2014:a} noted that the use of multiple
physiological signals is expected to enhance estimation of mental workload 
if the chosen signals represent separate aspects of
workload. They extracted features from EEG, ECG, skin conductance,
respiration, pupil size, and eye blinks, using these
as inputs for both support vector machine (SVM) and elastic net classifiers. They
achieved a high classification accuracy in two-level workload
determination.

Cognitive state can be estimated also through a combination of a single physiological signal
and a non-physiological source -- \citet{muhl:2014:a}, for example, used an LDA classifier to
classify mental workload, combining EEG and mood
(measured via self-reporting). Fusing of multiple features from the
same signal has been undertaken, for instance,
by \citet{brouwer:2012:a}, who used features derived from EEG as
inputs to an SVM classifier. They concluded
that models using both EEG and ERP features -- i.e., fusion models --
worked better for mental workload classification using short data
segments.

Given that learning can be facilitated by an optimal level of arousal \citep{Baldi2005,Sage1973},
\citet{Cowley2014} showed that the interaction of EDA and EEG features predicted learning outcomes
in a game-like task wherein individual signals alone were uninformative.

A study by \citet{wilson:2003:a} applied a combination of 85 EEG frequency power
features, inter-beat interval, respiration rate, and
electro-oculogram data from air traffic controllers. These 88
features were classified via an artificial neural network (ANN),
achieving very high accuracy in two-class mental workload
classification. Similar performance was achieved in classification of
mental workload on a three-level scale instead \citep{wilson:2003:b}. The same set of
signals were used by \citet{wilson:2007:a} in an
adaptive-aiding system.

\subsubsection{Applications}

Multimodal systems capable of interpreting physiological signals in realistic HCI
applications have been implemented with some success, especially in the last 10 years. We will discuss the most
relevant applied work here.

\citet{healey2005detecting} examined drivers' stress levels (referring to distress, or negative stress) via four modalities: ECG, EMG, EDA, and respiration. Using Fisher's discriminant analysis, they reported an accuracy of 97.4\%.
This relatively early research focused only on assessing the feasibility of correctly classifying three levels of stress. \citet{malta2011analysis} conducted a similar study, in which drivers' frustration was measured during actual (non-simulated) driving. Using a Bayesian network, they classified specific time segments in two classes (frustration-present vs. frustration-absent). They achieved, for the most part, 80\% hit and 9\% false positive rates, with the rates depending on the data fed to the classifier. Interestingly, the most accurate predictions were obtained when contextual information, such as traffic density, was provided to the classifier along with physiological information.

\citet{pavlidis2007computer} utilised thermal cameras to measure multiple signals. Although only a single sensor was employed technically (the camera's thermal sensor), they effectively measured blood flow, cardiac pulse, and breathing rate. These metrics were used to compute a general level of negative stress, by means of a model of data fusion developed specifically for this application. The researchers reported good accuracy in detecting the actual stress levels (\textit{r} = 0.91, Pearson correlation), with the exception of one outlier. They proposed two applications, desktop computer monitoring and sleep analysis, in both of which a stress level was computed for the current user, who was then alerted in the event that a specific threshold was exceeded. It should be noted, though, that the desktop computer test was carried out in the form of a laboratory experiment (a variant of the Stroop Test) \citep{pavlidis2007computer}.

In other work, a biometric mouse and finger sensor (intended for desktop computer users) were used to measure EDA, skin temperature, heart rate, and touch intensity, in conjunction with behavioural data (such as mouse movements), for development of a recommender system that was tested in real-world working
conditions over a span of four years \citep{kaklauskas2011web}. The recommender system was
capable of detecting stress and anxiety. It then presented individual users with suggestions for managing their work environment, on demand. Similar research involved a multimodal system employing EEG, ECG, and EMG, which was tested by CAD engineers. The system classified four emotions (frustration, challenge, engagement, and satisfaction) in users carrying out CAD work (so that, for example, one could detect designs developed amid frustration, which might display errors in judgement).
However, small sample sizes prevented statistically significant findings \citep{liu2013fuzzy}. In other applications, it was noted that less expert users might benefit from physiological integration, as in a system that utilises gaze, facial recognition, and speech recognition in order to predict user intentions. However, while welcomed by non-experts, a system of this nature \citep{maat2007gaze} was met with dissatisfaction by more experienced users.

Measurement of physiology from multiple sources has not been limited to single-user desktop computer usage. It has been carried out in more social settings also, such as at meetings or on public speaking occasions. For example, stress-related arousal has been measured `in the wild' through a combination of EDA, ECG, and motion sensors (though the system was tested with only a small number of participants) \citep{kusserow2013monitoring}. This creates the possibility of analysing one's own data during a public speaking event, so that any performance problems due to distress can be identified and, possibly, be avoided in the future. In addition, observer stress, the stress experienced by a person observing a meeting or group event, has been measured by means of thermal cameras, EDA, and EEG \citep{sharma2014modeling}. In the same study, observer stress was measured during an interview event (with the interview being designed to induce observer stress at specific moments). With genetic algorithms and support vector machines, the presence (or absence) of observer stress was detectable with an accuracy of 98\%.
In a similar context, a more laboratory-oriented setting (which used virtual humans as `presenters') utilised video cameras (for facial recognition), eye tracking, and audio signals to measure observer interest in a given topic \citep{schuller2009being}. That study demonstrated that a virtual agent was capable of detecting loss of the user's interest and of switching topic accordingly. Although the frequent change of topic might have had an adverse impact on user understanding, the system was capable of correctly detecting changes in user interest (with a cross-correlation value of 0.72, using support vector regression).

\subsubsection{Challenges}

As are all the other methods mentioned, multimodal methods are affected by high noise levels at the
receiving end. The `bleeding-edge' nature of the technologies employed implies that signal quality
can vary greatly across devices. Moreover, the interference typical of body-based measurements (such as
adverse effects of muscle movements on EEG) complicates processing of the signals obtained. In general, this means that `one size fits all' measurements are infeasible, so user- and
environment-based customisations are required for obtaining satisfactory signal quality,
especially when physiology is being measured outside well-controlled laboratory settings
\citep{hong2014affect}.

If they are to be effective, multimodal measurements would benefit from an overarching theory of emotion.
Although some approaches pointing in this direction have been put forward
\citep{conati2009empirically,peter2006emotion,gunes2013categorical,gunes2010automatic}, consensus is
still lacking, as the nature of emotions is not fully understood. Moreover, a lack
of interdisciplinary work is evident in this field: applications devised within one community are rarely
shared with other communities (for example, an HCI experiment may not demonstrate strong enough correlations between the data
and specific events to satisfy the demands of psychological research, while psychological research is usually
too theoretically oriented for HCI applications) \citep{lopatovska2011theories}.

The absence of consensus with regard to theories of emotion might also explain why multimodal methods generally
manifest only modest improvements over single-sensor settings \citep{dmello2012consistent}. Often, multimodal
systems assume co-ordinated bodily responses to a single emotion, which might not occur in
reality. Since multimodal measurements demand more computing power, it is important
to verify that multiple theories of emotion are considered while the work is still in the development stage, so as to
avoid the creation of a highly complex system that yields minimal gains over a single-sensor-based system.
That said, modest gains are still welcome, provided that the increase in computation time is
not unmanageable. Moreover, development of novel systems that take current theories of emotion into account could lead to 
new insights into the nature and physiology of emotions themselves (e.g., in the case of unexpected findings).

\subsubsection{Conclusion}

Most of the multimodal applications described here, especially in the context of HCI, have focused on detection of (negative) stress, possibly because stress
is often identified as the most negative -- yet measurable -- factor that can arise in working environments.
Applications in this area have found some success, although they rarely address use cases
beyond user self-assessment. Multimodal applications that constitute attempts to detect other cognitive or affective states,
such as interest, intention, or emotion, do exist but are only in the early stages of their 
development. For that reason, research in this connection is currently focused on improving classification accuracies rather than on
presenting effective use cases.

It appears that research into multimodal physiological systems would benefit from more interdisciplinary
work focused on finding common ground between the fields of psychology and human--computer interaction, so that 
advanced systems, capable of detecting user states with greater accuracy, can be developed.

\bibliographystyle{plainnat}
\bibliography{main}

\end{document}